# An ultrahot gas-giant exoplanet with a stratosphere

Thomas M. Evans[1], David K. Sing[1], Tiffany Kataria[2], Jayesh Goyal[1], Nikolay Nikolov[1], Hannah R. Wakeford[3], Drake Deming[4], Mark S. Marley[5], David S. Amundsen[6,7], Gilda E. Ballester[8], Joanna K. Barstow[9], Lotfi Ben-Jaffel[10], Vincent Bourrier[11], Lars A. Buchhave[12], Ofer Cohen[13], David Ehrenreich[11], Antonio García Muñoz[14], Gregory W. Henry[15], Heather Knutson[16], Panayotis Lavvas[17], Alain Lecavelier des Etangs[10], Nikole K. Lewis[18], Mercedes López-Morales[19], Avi M. Mandell[3], Jorge Sanz-Forcada[20], Pascal Tremblin[21] & Roxana Lupu[22]

[1]Astrophysics Group, School of Physics, University of Exeter, Stocker Road, Exeter EX4 4QL, UK. [2]NASA Jet Propulsion Laboratory, 4800 Oak Grove Drive, Pasadena, California 91109, USA. [3]NASA Goddard Space Flight Center, Greenbelt, Maryland 20771, USA. [4]Department of Astronomy, University of Maryland, College Park, Maryland 20742, USA. [5]NASA Ames Research Center, MS 245-5, Moffett Field, California 94035, USA. [6]Department of Applied Physics and Applied Mathematics, Columbia University, New York, New York 10025, USA. [7]NASA Goddard Institute for Space Studies, New York, New York 10025, USA. [8]Lunar and Planetary Laboratory, University of Arizona, Tucson, Arizona 85721, USA. [9]Department of Physics and Astronomy, University College London, Gower Street, London WC1E 6BT, UK. [10]Sorbonne Universités, UPMC Université Paris 6 and CNRS, UMR 7095, Institut d'Astrophysique de Paris, 98 bis boulevard Arago, F-75014 Paris, France. [11]Observatoire de l'Université de Genève, 51 chemin des Maillettes, 1290 Sauverny, Switzerland. [12]Centre for Star and Planet Formation, Niels Bohr Institute and Natural History Museum, University of Copenhagen, DK-1350 Copenhagen, Denmark. [13]Lowell Center for Space Science and Technology, University of Massachusetts, Lowell, Massachusetts 01854, USA. [14]Zentrum für Astronomie und Astrophysik, Technische Universität Berlin, D-10623 Berlin, Germany. [15]Center of Excellence in Information Systems, Tennessee State University, Nashville, Tennessee 37209, USA. [16]Division of Geological and Planetary Sciences, California Institute of Technology, Pasadena, California 91125, USA. [17]Groupe de Spectrométrie Moléculaire et Atmosphérique, UMR 7331, CNRS, Université de Reims Champagne-Ardenne, Reims 51687, France. [18]Space Telescope Science Institute, 3700 San Martin Drive, Baltimore, Maryland 21218, USA. [19]Harvard-Smithsonian Center for Astrophysics, 60 Garden Street, Cambridge, Massachusetts 02138, USA. [20]Centro de Astrobiología (CSIC-INTA), ESAC Campus, Camino bajo del Castillo, E-28692 Villanueva de la Cañada, Madrid, Spain. [21]Maison de la Simulation, CEA, CNRS, Université Paris-Sud, UVSQ, Université Paris-Saclay, 91191 Gif-sur-Yvette, France. [22]Bay Area Environmental Research Institute, Moffett Field, California 94035, USA.

**Infrared radiation emitted from a planet contains information about the chemical composition and vertical temperature profile of its atmosphere[1–3]. If upper layers are cooler than lower layers, molecular gases will produce absorption features in the planetary thermal spectrum[4,5]. Conversely, if there is a stratosphere—where temperature increases with altitude—these molecular features will be observed in emission[6–8]. It has been suggested that stratospheres could form in highly irradiated exoplanets[9,10], but the extent to which this occurs is unresolved both theoretically[11,12] and observationally[3,13–15]. A previous claim for the presence of a stratosphere[14] remains open to question, owing to the challenges posed by the highly variable host star and the low spectral resolution of the measurements[3]. Here we report a near-infrared thermal spectrum for the ultrahot gas giant WASP-121b, which has an equilibrium temperature of approximately 2,500 kelvin. Water is resolved in emission, providing a detection of an exoplanet stratosphere at $5\sigma$ confidence. These observations imply that a substantial fraction of incident stellar radiation is retained at high altitudes in the atmosphere, possibly by absorbing chemical species such as gaseous vanadium oxide and titanium oxide.**

We observed a secondary eclipse of WASP-121b on 10 November 2016 using the Hubble Space Telescope (HST) Wide Field Camera 3 (WFC3). Time series spectra were acquired using the G141 grism, which covers the 1.1–1.7 μm wavelength range. A secondary eclipse of WASP-121b was also observed with the Spitzer space telescope on 30 January 2017 in the 3.6 μm photometric channel of the Infrared Array Camera (IRAC). Further details of the observing set-up and reduction of these data sets are provided in Methods.

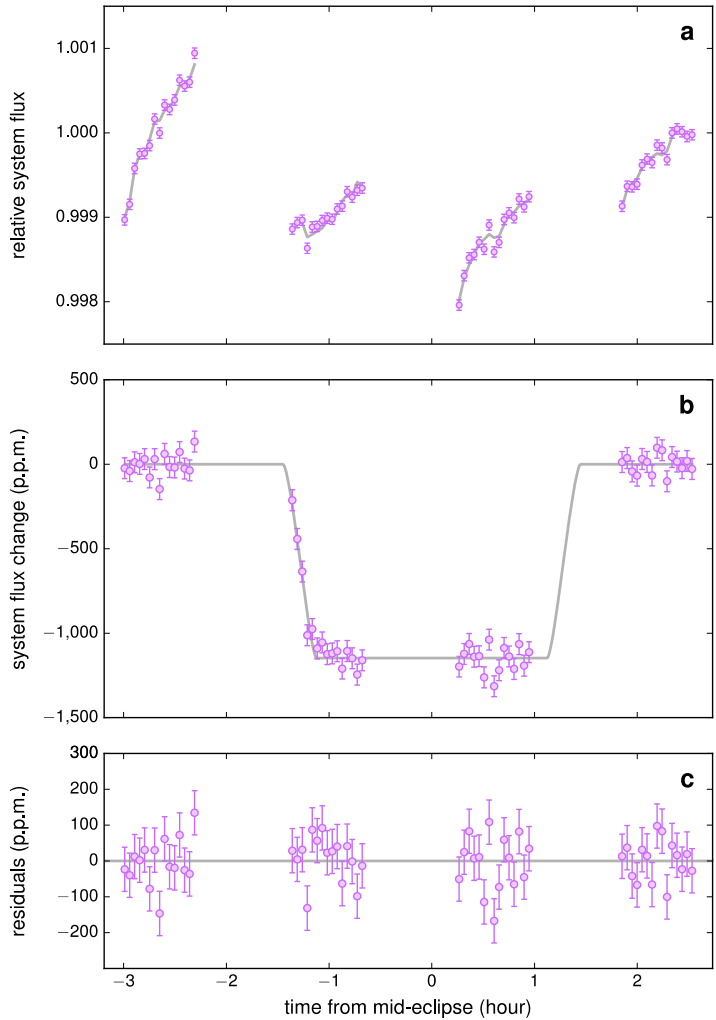

**Figure 1 | Wavelength-integrated white light curve for WASP-121b. a**, Raw normalized flux with photon-noise $1\sigma$ error bars (pink circles) and best-fit Gaussian process model (grey line). Gaps in the time series are due to the target being occulted by the Earth during each HST orbit. Quasi-repeatable systematics are evident for each HST orbit as well as a visit-long drift in the baseline flux level. **b**, Relative change in the system flux in parts per million (p.p.m.) with photon-noise $1\sigma$ error bars (pink circles) after correcting for the instrumental systematics, with best-fit eclipse model (grey line). **c**, Best-fit model residuals in p.p.m. with photon-noise $1\sigma$ error bars (pink circles).

For the WFC3 data, we produced a wavelength-integrated 'white' light curve by summing each spectrum along the dispersion axis (Fig. 1). The white light curve exhibits instrumental systematics, most notably a quasi-repeatable ramp within each HST orbit, which is a well-known systematic that affects HST/WFC3 data[4,5,16–19]. We modelled the planet eclipse signal and instrumental systematics simultaneously by treating the data as a Gaussian process. The eclipse mid-time $T_{mid}$ and planet-to-star surface flux ratio $F_p/F_s$ were allowed to vary in the fitting, while the planet-to-star radius ratio $R_p/R_s$, normalized semi-major axis $a/R_s$, orbital inclination $i$, and orbital period $P$ were held fixed to previously determined values[19,20]. Details of our light curve model and fitting procedure for both WFC3 and IRAC are provided in Methods. The best-fit model for the WFC3 light curve is shown in Fig. 1, and results are reported in Extended Data Table 1.

We produced spectroscopic light curves by summing each WFC3 spectrum within 28 channels across the 1.122–1.642 μm wavelength range. As described in Methods, we adopted a commonly used approach to remove wavelength-independent systematics as well as systematics arising from pointing shifts along the dispersion axis for each spectroscopic light curve[19]. The resulting time series are well-behaved and exhibit minimal residual correlations (Extended Data Fig. 1). We fitted the spectroscopic light curves using the same Gaussian process methodology as described above and in Methods for the white light curve analysis, but only allowed $F_p/F_s$ to vary while holding $T_{mid}$ fixed to the value determined for the white light curve. The best-fit models are shown in Extended Data Fig. 1 and the results are reported in Extended Data Table 2.

The wavelength-dependent eclipse depths measured with HST and Spitzer are shown in Fig. 2 and comprise the thermal emission spectrum for WASP-121b. A number of previous measurements of thermal emission from ultrahot gas giants have been consistent with radiation from isothermal blackbodies[16,18]. This may indicate that the atmospheric temperature remains constant over the pressures probed by the different wavelengths, or that the same pressure level is probed at all wavelengths, which could be the case if there is an optically thick cloud deck across the dayside hemisphere. Alternatively, the data precision may simply not be high enough to exclude an isothermal blackbody. When we fitted an isothermal blackbody to the WASP-121b thermal spectrum, we obtain a best-fit atmospheric temperature of $2,700 \pm 10$ K. However, this model, which is shown in Fig. 2 by the yellow line, provides a poor fit to the data with a $\chi^2$ of 83.9 for 29 degrees of freedom. This allows us to exclude an isothermal blackbody spectrum for WASP-121b at $5\sigma$ confidence.

The inability of an isothermal blackbody to explain the data implies that the thermal spectrum probes a range of atmospheric layers with different temperatures. Given that the emission is strongest at 1.35–1.55 μm and 1.20–1.25 μm (Fig. 2), the temperature must be higher at pressures for which the atmosphere is optically thick at these wavelengths. For a decreasing temperature-pressure ($T$–$P$) profile (that is, where temperature decreases as pressure decreases), this would only be possible if the opacity of the atmosphere were relatively low within these bands compared to the surrounding wavelengths. However, no plausible gas-phase absorber—such as $H_2O$, CO, VO, TiO, FeH or CrH—has this property[21] (see Methods and Extended Data Fig. 2), and refractory condensates (such as $Al_2O_3$, $CaTiO_3$, FeO, $MgSiO_3$ and $MgSiO_4$) are spectrally featureless within the WFC3 bandpass[22]. An increasing $T$–$P$ profile is the only remaining possibility, with the strength of the thermal emission positively correlating with the atmospheric opacity.

To further interpret the measured spectrum, we performed a retrieval analysis to constrain the $T$–$P$ profile and chemical species present in the atmosphere[1–3]. In our retrieval, we allowed the abundances of $H_2O$, TiO, VO, FeH, CrH, CO, $CH_4$ and $NH_3$ to vary as free parameters, and assumed that these gases are well mixed vertically throughout the atmosphere. A one-dimensional analytic formulation was adopted for the $T$–$P$ profile, which assumes radiative equilibrium and is flexible enough to describe atmospheres with and without stratospheres[2,3,23]. Further details of our retrieval methodology are provided in Methods.

Our best-fit thermal spectrum from the retrieval analysis is shown by the red line in Fig. 2, and gives a good fit to the data with a $\chi^2$ value of 26.6 for 25 degrees of freedom. This model includes only $H_2O$ and VO opacity, as we found the inclusion of other molecules was not justified by the current data (see Methods). The retrieved $T$–$P$ profiles are shown in Fig. 3, all of which exhibit a stratosphere between pressures of $10^{-1}$ bar and $10^{-5}$ bar, with an inferred temperature increase of $1,114^{+330}_{-313}$ K across this pressure range.

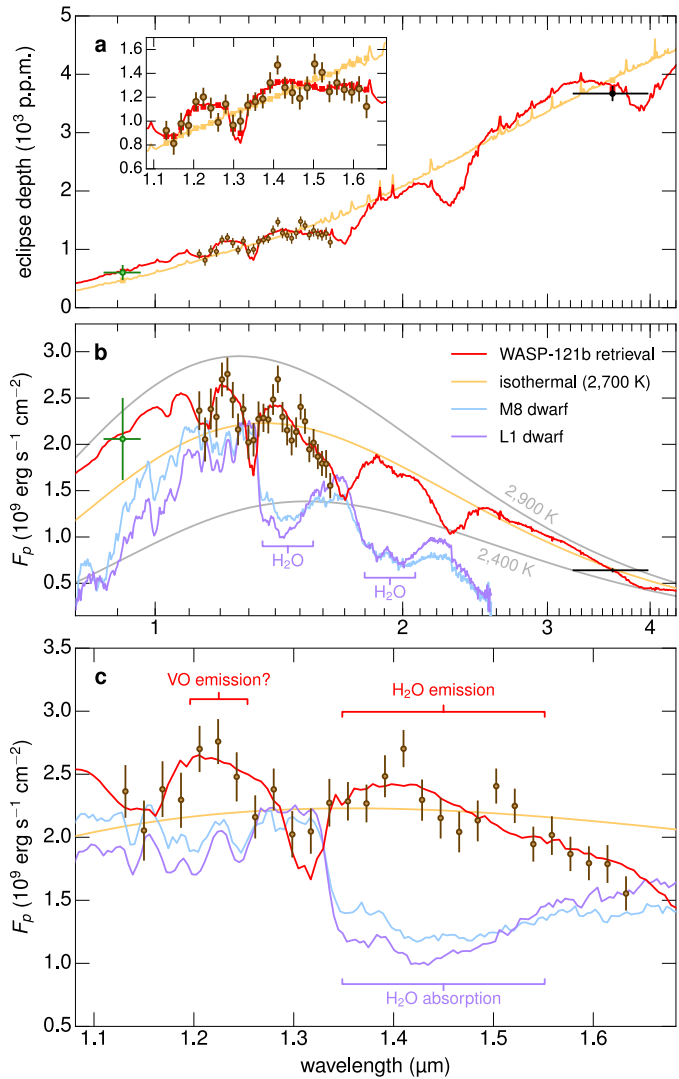

**Figure 2 | Emission spectrum for WASP-121b. a,** Eclipse depths measured with HST (brown circles) and Spitzer (black circle, top right), and a ground-based photometric measurement[20] (green circle). Vertical error bars give $1\sigma$ uncertainties and horizontal error bars give photometric bandpasses. Yellow line shows the best-fit isothermal blackbody spectrum with a temperature of 2,700 K. Red line shows the best-fit model from the retrieval analysis. Inset shows the HST data on a magnified scale. **b,** Planetary flux with stellar contribution removed. Measured spectra for M8 and L1 dwarfs[30] are shown for comparison (blue and purple lines), exhibiting $H_2O$ absorption bands (bracketed) characteristic of these spectral types. For WASP-121b the $H_2O$ band at 1.35–1.55 μm appears in emission. Spectra for 2,400 K and 2,900 K blackbodies (grey lines) indicate the approximate temperature range probed by the data. **c,** Similar to **b**, but showing only the WFC3 bandpass and with a possible emission feature due to VO also indicated at 1.20–1.25 μm.

Detection of the stratosphere is driven by the emission peaks at 1.35–1.55 μm and 1.20–1.25 μm, which coincide with $H_2O$ and VO bands (see Fig. 2 and Extended Data Figs 2 and 3). Indeed, $H_2O$ has previously been detected in the transmission spectrum of WASP-121b[19], and high-altitude absorption of incident stellar radiation at optical wavelengths by VO, as well as by TiO, has been proposed as a possible mechanism for forming stratospheres in strongly irradiated exoplanets[9,10]. However, although the inferred $H_2O$ abundance is broadly consistent with solar elemental abundances in chemical equilibrium, the 95% lower credible limit for the inferred VO abundance is about 1,000× solar. Such a high inferred VO abundance warrants scepticism. For instance, there is a well-known modelling degeneracy between the absolute pressure level of a stratosphere and the abundance of optical absorbers such as VO, with higher abundances giving stratospheres at lower pressures and vice versa[10]. It is also possible that the relatively simple $T$–$P$ profile adopted in the retrieval analysis produces a biased VO abundance estimate, as it forces the atmosphere to be isothermal at low pressures (Fig. 3) and does not account for three-dimensional effects or non-equilibrium chemistry. Alternatively, the 1.20–1.25 μm feature could be emission from some other unknown source. Given these caveats, a robust detection or non-detection of VO must await further observations that are capable of spectrally resolving the emission bands at high signal-to-noise ratio; for example, with the upcoming James Webb Space Telescope. Nonetheless, the conclusion that WASP-121b has a stratosphere appears secure, as the data resolve $H_2O$ in emission while ruling out isothermal and decreasing *T–P* profiles at high confidence.

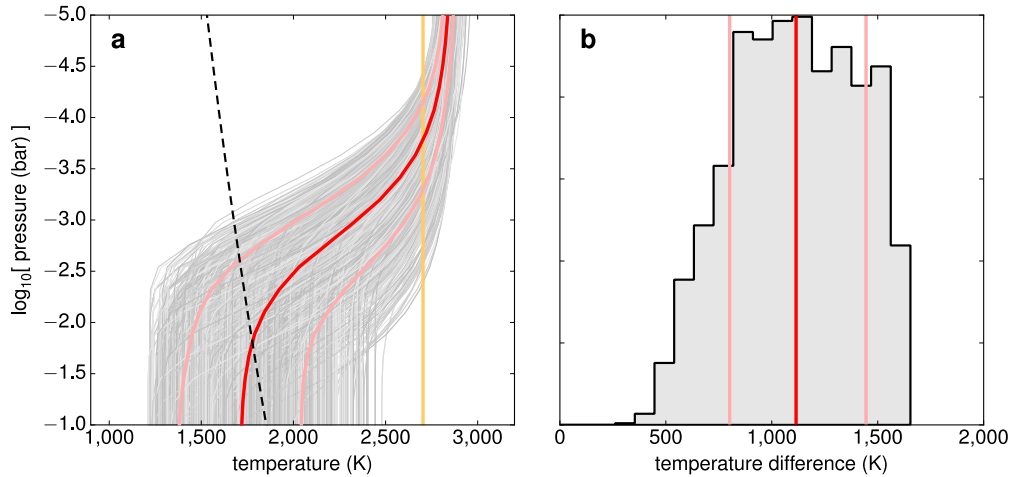

**Figure 3 | Temperature–pressure profiles for WASP-121b. a**, Grey lines show a random subset of *T–P* profiles sampled by the MCMC retrieval analysis. Red line shows the median temperature at each pressure level, and pink lines show ranges either side encompassing ±34% of the sampled profiles. Yellow line indicates the best-fit isothermal temperature of 2,700 K. Dashed line shows the approximate condensation curve for Ti and V compounds[28]. **b**, Histogram showing the temperature difference between pressures of $10^{-1}$ bar and $10^{-5}$ bar. The median increase across this pressure range is indicated by the red line, with pink lines showing the ranges either side encompassing ±34% of the samples.

Using an approach described in previous studies[9,24], we calculate that a heat deposit equivalent to more than 20% of the incident stellar flux at pressures below $10^{-3}$ bar is required to raise the stratosphere temperature to about 2,700 K (Fig. 3). Direct absorption of impinging stellar radiation by gaseous species with strong optical bands seems the most plausible means to achieve this. To estimate the required opacity of the putative absorber, we generated radiative-convective equilibrium models including a parameterized grey absorber across optical wavelengths[11,12] assuming 1× and 15× solar elemental abundances, with TiO and VO opacities removed. We find the predicted emission spectra and $H_2O$ emission features are in reasonable agreement with the data for opacities of 0.02 $cm^2 g^{-1}$ for the 1× solar case and 0.06 $cm^2 g^{-1}$ for the 15× solar case (Extended Data Fig. 4). Alternative heating mechanisms, such as the breaking of upward-propagating gravity waves, are not expected to produce energy fluxes of the required magnitude[25,26]. Strong absorption within high-altitude clouds could also potentially produce a stratosphere[27]. However, it is unclear whether a significant cloud deck could be maintained on the dayside hemisphere of WASP-121b, as the high temperatures are likely to inhibit the condensation of even the most refractory compounds of elements such as V, Ti, Ca and Fe (ref. 28).

WASP-121b is the first exoplanet with a stratosphere for which resolved spectral features have been observed in emission. Previous stratosphere detection claims have either been revised[3,13] or are based on measurements of excess thermal emission without resolved spectral features[14,15]. Published thermal spectra for other gas-giant exoplanets are either indistinguishable from isothermal emission otherwise consistent with decreasing temperature profiles[4,5,17]. These results are in line with the overall finding that irradiated gas-giant atmospheres exhibit a wide range of properties[29]. Expanding the sample of ultrahot exoplanets with measured thermal emission spectra will be central to making sense of this diversity, particularly as we enter the era of the James Webb Space Telescope, which is anticipated to vastly improve the quality of exoplanet spectra across the near-to-mid-infrared wavelength range.

1. Madhusudhan, N. & Seager, S. A temperature and abundance retrieval method for exoplanet atmospheres. *Astrophys. J.* **707,** 24–39 (2009).
2. Line, M. R. *et al.* A systematic retrieval analysis of secondary eclipse spectra. I. A comparison of atmospheric retrieval techniques. *Astrophys. J.* **775,** 137 (2013).
3. Line, M. R. *et al.* No thermal inversion and a solar water abundance for the hot Jupiter HD 209458b from HST/WFC3 spectroscopy. *Astrophys. J.* **152,** 203 (2016).
4. Stevenson, K. B. *et al.* Thermal structure of an exoplanet atmosphere from phase-resolved emission spectroscopy. *Science* **346,** 838–841 (2014).
5. Beatty, T. G. *et al.* Evidence for atmospheric cold-trap process in the non-inverted emission spectrum of Kepler-13Ab using HST/WFC3. Preprint at https://arxiv.org/abs/1612.06409 (2016).
6. Gillett, F. C., Low, F. J. & Stein, W. A. The 2.8–14-micron spectrum of Jupiter. *Astrophys. J.* **157,** 925–934 (1969).
7. Ridgway, S. T. Jupiter: identification of ethane and acetylene. *Astrophys. J.* **187,** L41–L43 (1974).
8. Wallace, L., Prather, M. & Belton, M. J. S. The thermal structure of the atmosphere of Jupiter. *Astrophys. J.* **193,** 481–493 (1974).
9. Fortney, J. J., Lodders, K., Marley, M. S. & Freedman, R. S. A unified theory for the atmospheres of the hot and very hot Jupiters: two classes of irradiated atmospheres. *Astrophys. J.* **678,** 1419–1435 (2008).
10. Hubeny, I., Burrows, A. & Sudarsky, D. A possible bifurcation in atmospheres of strongly irradiated stars and planets. *Astrophys. J.* **594,** 1011–1018 (2003).
11. Burrows, A., Budaj, J. & Hubeny, I. Theoretical spectra and light curves of close-in extrasolar giant planets and comparison with data. *Astrophys. J.* **678,** 1436–1457 (2008).
12. Spiegel, D. S., Silverio, K. & Burrows, A. Can TiO explain thermal inversions in the upper atmospheres of irradiated giant planets? *Astrophys. J.* **699,** 1487–1500 (2009).
13. Knutson, H. A., Charbonneau, D., Allen, L. A., Burrows, A. & Megeath, S. T. The 3.6–8.0μm broadband emission spectrum of HD209458b: evidence for an atmospheric temperature inversion. *Astrophys. J.* **673,** 526–531 (2008).
14. Haynes, K., Mandell, A. M., Madhusudhan, N., Deming, D. & Knutson, H. Spectroscopic evidence for a temperature inversion in the dayside atmosphere of hot Jupiter WASP-33b. *Astrophys. J.* **806,** 146 (2015).
15. Wong, I. *et al.* 3.6 and 4.5 μm Spitzer phase curves of the highly irradiated hot Jupiters WASP-19b and HAT-P-7b. *Astrophys. J.* **823,** 122 (2016).
16. Cartier, K. M. S. *et al.* Near-infrared emission spectrum of WASP-103b using Hubble Space Telescope/Wide Field Camera 3. *Astron. J.* **153,** 34 (2017).
17. Crouzet, N., McCullough, P. R., Deming, D. & Madhusudhan, N. Water vapor in the spectrum of the extrasolar planet HD 189733b. II. The eclipse. *Astrophys. J.* **795,** 166 (2014).
18. Wilkins, A. N. *et al.* The emergent 1.1–1.7 μm spectrum of the exoplanet COROT-2b as measured using the Hubble Space Telescope. *Astrophys. J.* **783,** 113 (2014).
19. Evans, T. M. *et al.* Detection of $H_2O$ and evidence for TiO/VO in an ultra-hot exoplanet atmosphere. *Astrophys. J.* **822,** L4 (2016).
20. Delrez, L. *et al.* WASP-121 b: a hot Jupiter close to tidal disruption transiting an active F star. *Mon. Not. R. Astron. Soc.* **458,** 4025–4043 (2016).
21. Sharp, C. M. & Burrows, A. Atomic and molecular opacities for brown dwarf and giant planet atmospheres. *Astrophys. J. Suppl. Ser.* **168,** 140–166 (2007).
22. Wakeford, H. R. & Sing, D. K. Transmission spectral properties of clouds for hot Jupiter exoplanets. *Astron. Astrophys.* **573,** A122 (2015).
23. Guillot, T. On the radiative equilibrium of irradiated planetary atmospheres. *Astron. Astrophys.* **520,** A27 (2010).
24. Marley, M. S. & McKay, C. P. Thermal structure of Uranus' atmosphere. *Icarus* **138,** 268–286 (1999).
25. Hickey, M. P., Walterscheid, R. L. & Schubert, G. Gravity wave heating and cooling in Jupiter's thermosphere. *Icarus* **148,** 266–281 (2000).
26. Matcheva, K. I. & Strobel, D. F. Heating of Jupiter's thermosphere by dissipation of gravity waves due to molecular viscosity and heat conduction. *Icarus* **140,** 328–340 (1999).
27. Heng, K., Hayek, W., Pont, F. & Sing, D. K. On the effects of clouds and hazes in the atmospheres of hot Jupiters: semi-analytical temperature-pressure profiles. *Mon. Not. R. Astron. Soc.* **420,** 20–36 (2012).
28. Lodders, K. Titanium and vanadium chemistry in low-mass dwarf stars. *Astrophys. J.* **577,** 974–985 (2002).

29. Sing, D. K. *et al.* A continuum from clear to cloudy hot-Jupiter exoplanets without primordial water depletion. *Nature* **529,** 59–62 (2016).

30. Kirkpatrick, J. D. *et al.* Discoveries from a near-infrared proper motion survey using multi-epoch two micron all-sky survey data. *Astrophys. J. Suppl. Ser.* **190,** 100–146 (2010).

**Acknowledgements** This work is based on observations with the NASA/ESA HST, obtained at the Space Telescope Science Institute (STScI) operated by AURA, Inc. This work is also based in part on observations made with the Spitzer Space Telescope, which is operated by the Jet Propulsion Laboratory, California Institute of Technology under a contract with NASA. The research leading to these results has received funding from the European Research Council under

**Author Contributions** T.M.E. and D.K.S. designed the HST observations of WASP-121. D.K.S. and M.L.-M. led the HST Treasury programme, with support provided by all authors. T.M.E. led the HST data analysis with contributions from N.N., H.R.W. and D.D. D.D. proposed and designed the Spitzer observations and analysed the data. D.K.S. led the retrieval analysis. T.K., J.G., M.S.M., A.L.E. and P.T. provided additional theoretical interpretation of the data. R.L. provided molecular absorption cross-sections for the theoretical interpretation. T.M.E. wrote the manuscript along with D.K.S., T.K., M.S.M. and A.L.E. All authors discussed the results and commented on the paper. The author list ordering is alphabetical after M.S.M.

# METHODS

## HST/WFC3 observations

A secondary eclipse of WASP-121b was observed using HST/WFC3 as part of Program 14767 (PIs D.K.S. and M.L.-M.). We observed the target for 6.9 h over five HST orbits, which covered the full planetary eclipse lasting 2.9 h. The first two HST orbits occurred before eclipse ingress, the third and fourth orbits occurred during the eclipse, and the fifth orbit occurred after the completion of eclipse egress (Fig. 1).

Observations were made in spectroscopic mode using forward-scanning with the G141 grism and a scan rate of 0.12 arcsec $s^{-1}$. Overheads were reduced by only reading out the $256 \times 256$ pixel subarray containing the target spectrum. We adopted the SPARS10 sampling sequence with 15 non-destructive reads per exposure (NSAMP = 15) resulting in total integration times of 103 s and scans across 100-pixel rows of the cross-dispersion axis. With this configuration, we obtained 14 science exposures in the first HST orbit and 16 exposures in each subsequent HST orbit. Typical count levels remained below $2.5 \times 10^4$ electrons per pixel.

## HST/WFC3 data reduction

The target flux was extracted from the IMA files produced by the CALWF3 pipeline (v3.4) by taking the difference between successive non-destructive reads. The background was removed by subtracting the median count in a box of pixels spanning 110 columns along the dispersion axis and 20 rows along the cross-dispersion axis well away from the target spectrum. For each read-difference, we determined the flux-weighted centre of the scanned spectrum along the cross-dispersion axis and set to zero all pixel values located more than 35 pixels above and below this pixel row. This removed flux contributions from nearby contaminant stars and cosmic ray strikes from the masked region. Final reconstructed frames were produced by adding together the read-differences.

The target spectrum was then extracted from each frame by summing the flux within a rectangular aperture spanning the full dispersion axis and positioned on the central cross-dispersion row of the scan. We did this for a range of different cross-dispersion aperture widths spanning 100 to 200 pixels in ten-pixel increments. The wavelength solution was determined by cross-correlating each target spectrum against a Phoenix stellar model[31] with properties similar to the WASP-121 host star (effective temperature $T_{eff}$ = 6,400 K, log[$g$ (cgs)] = 4.0, Fe/H = 0) modulated by the throughput of the G141 grism.

## HST/WFC3 white light curve analysis

White light curves were generated for each trial aperture by summing the spectra along the dispersion axis (Fig. 1). Before fitting the light curves, we discarded the first HST orbit because of the particularly strong ramp systematic it exhibited, which is a well-known feature of WFC3 data sets[4,5,16–19,32,33]. We also discarded the first exposure of each remaining HST orbit as these had significantly lower counts than subsequent exposures in the same orbit.

To model the white light curves we adopted a Gaussian process methodology similar to that outlined in our previous work[19]. We defined our data likelihood as a multivariate normal distribution with a deterministic mean function $\mu$ to model the planetary eclipse signal and covariance matrix $K$ to capture additional correlations in the data, such that:

$$p(d \mid u, w) = N(\mu, K + \Sigma)$$

where $p$ denotes a probability density function, $d$ is a vector containing the flux measurements, $u$ is a vector containing the mean function parameters, $w$ is a vector containing the covariance parameters, $N$ denotes a multivariate normal distribution, and $\Sigma$ is a diagonal matrix containing the squared photon-noise uncertainties for each data point. For the mean function $\mu$, we used:

$$\mu(t, t'; c_0, c_1, F_p/F_s, T_{mid}) = [c_0 + c_1 \, t'] \, E(t; F_p/F_s, T_{mid})$$

where $t$ is a vector containing the observation timestamps, $t'$ is standardized time (that is, time $t$ minus the mean divided by the standard deviation), $c_0$ and $c_1$ are parameters defining a linear trend, $E$ is the analytic function for a planetary eclipse taken from ref. 34, $F_p/F_s$ is the planet-to-star flux ratio, and $T_{mid}$ is the eclipse mid-time. For the eclipse signal $E$ we ignored limb darkening and fixed the remaining parameters to the previously determined values listed in Extended Data Table 1. For the covariance function, we adopted a Matérn $\nu = 3/2$ kernel with HST orbital phase $\varphi$, dispersion drift $x$, and cross-dispersion drift $y$ as the input variables. As with time $t$ for the linear trend in the mean function $\mu$, each of these input variables was standardized before being provided as input to the Gaussian process (that is, mean-subtracted and divided by the standard deviation). Thus, the entries of the covariance matrix were given by:

$$K_{ij} = A^2 \, (1 + D_{ij} \, 3^{1/2}) \exp[-D_{ij} \, 3^{1/2}]$$

where $A$ defines the characteristic correlation amplitude and:

$$D_{ij} = [\eta^2_\varphi \, (\varphi'_i - \varphi'_j)^2 + \eta^2_x \, (x'_i - x'_j)^2 + \eta^2_y \, (y'_i - y'_j)^2]^{1/2}$$

where $\eta_\varphi$, $\eta_x$ and $\eta_y$ are the inverse correlation length scales and the primed variables are standardized. The free parameters of our white light curve model were therefore $u = [c_0, c_1, F_p/F_s, T_{mid}]$ for the mean function and $w = [A, \eta_\varphi, \eta_x, \eta_y]$ for the covariance function. Our prior distribution over the model parameters took the form $p(u, w) = p(c_0) \, p(c_1) \, p(F_p/F_s) \, p(T_{mid}) \, p(A) \, p(\eta_\varphi) \, p(\eta_x) \, p(\eta_y)$. Uniform distributions were adopted for $p(c_0)$, $p(c_1)$, $p(F_p/F_s)$, and $p(T_{mid})$. A gamma distribution was adopted for $p(A)$ with the form $\mathrm{Gamma}(\alpha=1, \beta=100) = 100 \exp[-100A]$. We found that priors less strongly favouring small values of $A$ (for example, a uniform or log-uniform distribution) resulted in the Gaussian process model being biased by a subset of outliers in the data to infer implausibly short correlation length scales with large correlation amplitudes. Effectively, this resulted in a model that was over-fitted to the data, justifying the adoption of the stronger gamma distribution prior. Log-uniform distributions were adopted for $p(\eta_\varphi)$, $p(\eta_x)$, and $p(\eta_y)$ (that is, we fitted for $\log[\eta_\varphi]$, $\log[\eta_x]$ and $\log[\eta_y]$ with uniform distribution priors).

The posterior distribution $p(u, w \mid d) \propto p(d \mid u, w) \, p(u, w)$ was marginalized using affine-invariant Markov chain Monte Carlo (MCMC) as implemented by the emcee Python software package[35].

We randomly initialized five groups of 150 walkers close to the maximum likelihood solution, which was located by minimizing the negative logarithm of the posterior distribution using nonlinear optimization as implemented by the fmin routine of the scipy.optimize Python software package[36]. Each of the five walker groups was run for 500 steps before discarding the first 250 steps as burn-in. The resulting five chains displayed good mixing and convergence, with Gelman-Rubin statistic values well within 1% for each model parameter[37]. The five individual chains were then combined to produce the final posterior samples.

This analysis was repeated for each of the trial apertures used to extract the target spectra (see above). All inferred values for $F_p/F_s$ agreed to within $1\sigma$ regardless of the aperture used. We adopt the 160-pixel aperture reduction for the remainder of this study and report the MCMC results in Extended Data Table 1. The standard deviation of the best-fit model residuals for this analysis was 64 p.p.m., which is within 4% of the theoretical photon-noise floor (Fig. 1), while the other apertures gave either larger uncertainties for $F_p/F_s$ or larger residual scatter relative to photon noise.

### HST/WFC3 spectroscopic light curve analysis

Spectroscopic light curves were produced using the same method as adopted in our previous work[19,32,33]. We started by identifying the out-of-eclipse spectra based on the best-fit white light curve model and taking the median of these to form a master stellar spectrum. For each individual spectrum, we determined the lateral shift along the dispersion axis and a wavelength-independent rescaling of the flux that minimized the residuals with the master spectrum. The residuals obtained in this way were binned into 28 spectroscopic channels across the 1.122–1.642 μm wavelength range, which avoided the steep edges of the G141 grism response. Each channel spanned 4 pixel columns along the dispersion axis, equivalent to 0.019 μm in wavelength. The time series of binned residuals were then added to the eclipse signal of the best-fit white light curve model to produce the spectroscopic light curves shown in Extended Data Fig. 1. This process removed systematics due to pointing drifts along the dispersion axis and the wavelength-independent ('common-mode') component of the flux time series. The latter included the ramp systematic, but also the wavelength-integrated eclipse signal, hence manually adding the eclipse signal from the white light curve best-fit model back in the final step.

To fit the spectroscopic light curves, we adopted the same data likelihood and prior distributions as were used for the white light curve analysis. The only exception was that we did not fit for $T_{mid}$ as a free parameter but instead held it fixed to the best-fit value determined for the white light curve. We also adopted the same fitting methodology that was used for the white light curve analysis, first identifying the maximum likelihood solution by nonlinear least squares and using it as a starting point for affine-invariant MCMC to marginalize over the posterior distribution. The results are reported in Extended Data Table 2 and best-fit models are shown in Extended Data Fig. 1. Median scatter in the best-fit model residuals across the 28 channels was equal to photon noise, with a $1\sigma$ range about the median of 90%–110% times photon noise.

As an aside, we note that the apparent size of WASP-121b when viewed from the zenith geometry at secondary eclipse could vary as a function of wavelength due to the wavelength-dependent opacity of the atmosphere. In transmission, this effect produces a variation in $R_p/R_s$ of up to ~0.4% about the mean across the WFC3 G141 bandpass[19]. Given that the eclipse depth scales as $(R_p/R_s)^2$, this translates to a potential bias in the inferred planetary flux of up to ~0.8% in any given wavelength channel, which is well within the $1\sigma$ uncertainties obtained from our light curve fitting (Extended Data Table 2).

**Spitzer/IRAC observations, data reduction and light curve analysis**

A single secondary eclipse of WASP-121b was observed using Spitzer/IRAC as part of Program 13044 (PI D.D.). Observations were made in the 3.6 μm photometric channel using a $32 \times 32$ pixel subarray and an exposure time of 2.0 s per image. We observed the target for 8.5 h, which included the full eclipse, as well as a pre-eclipse baseline of 3.2 h and a post-eclipse baseline of 2.4 h. Data reduction was performed using the methodology described in our previous work[38,39]. We analysed the light curve by fitting for an eclipse signal with $F_p/F_s$ and $T_{mid}$ allowed to vary, while treating instrumental systematics caused by the intrapixel sensitivity variations of the IRAC detector with pixel level decorrelation (PLD) as described in ref. 39. An 800,000 step MCMC was used to derive estimates for the model parameters, giving an eclipse depth of $(R_p/R_s)^2 F_p/F_s = 3{,}670 \pm 130$ p.p.m. and mid-time of $T_{mid} = 2{,}45$ $7{,}783.77740 \pm 0.00068$ (BJD$_{TDB}$).

**1D atmosphere modelling of WASP-121b**

We used the 1D atmosphere ATMO model[40–44] to perform forward and retrieval modelling of the measured WASP-121b thermal emission spectrum. ATMO computes the 1D temperature–pressure ($T$–$P$) structure of an atmosphere in plane-parallel geometry. It can calculate forward models assuming radiative, convective and chemical equilibrium. It can also be used as a retrieval tool to compute the emission and transmission spectra from an input $T$–$P$ profile and arbitrary chemical abundances[45].

For the retrieval analysis, we used a model planetary atmosphere with 50 pressure levels evenly spaced in log pressure between $10^{-8}$ bar and 500 bar. We adopted the parameterized $T$–$P$ profile of ref. 23, fitting the data assuming either one or two optical channels. This gave three to five parameters for the $T$–$P$ profile: the Planck mean thermal infrared opacity, $\kappa_{IR}$; the ratios of the optical to infrared opacities in the two channels, $\gamma_1$ and $\gamma_2$; a partition of the flux between the two optical channels, $\alpha$; and an irradiation efficiency factor, $\beta$. We set the planetary radius equal to 1.694 Jupiter radii ($R_J$) which corresponds to the lowest-altitude probed by the near-infrared transmission spectrum[19]. An internal planetary temperature of 100 K was assumed and the same Phoenix stellar model as described above was used for the input flux from the WASP-121 host star.

We adopted uniform priors for all free parameters in our model with the following ranges: $10^{-5}$ to $10^{-0.5}$ for $\kappa_{IR}$; $10^{-4}$ to $10^{1.5}$ for $\gamma_1$ and $\gamma_2$; 0 to 1 for $\alpha$; and 0 to 2 for $\beta$. Uniform priors between 0 and 0.05 were also adopted for the mixing ratios of chemical species other than H and He. The upper limit placed on the metal abundances is justified by the fact that WASP-121b is known to be a gas giant. We ran a suite of retrievals including various combinations of $H_2O$, TiO, VO, FeH, CrH, CO, $CH_4$ and $NH_3$, using nonlinear least-squares optimization to locate the minimum $\chi^2$ solution for each model. However, we found that the inclusion of molecules other than $H_2O$ and VO did not offer a significant improvement in the fit to the data. This is quantified in Extended Data Table 3, where we provide the full list of retrievals performed with accompanying Bayesian information criterion (BIC) values. We also found that using two optical channels rather than one for the $T$–$P$ profile did not improve the quality of the fit nor did it substantially alter the inferred stratosphere profile. Thus, our final model had five free parameters: volume mixing ratios for $H_2O$ and VO plus $\kappa_{IR}$, $\gamma_1$, and $\beta$. For this model, we marginalized over the posterior distribution using differential-evolution MCMC[46]. We ran 10 chains for 30,000 steps each, at which point the Gelman-Rubin statistic for each free parameter was within 1% of unity, indicating that the chains were well-mixed and had reached a steady state. Following ref. 46, we then discarded a burn-in phase from all chains corresponding to the step at which all chains had found a $\chi^2$ below the median $\chi^2$ value of the chain. The remaining samples were then combined into a single chain to estimate the posterior distribution. The results are reported in Extended Data Table 4 and plotted in Extended Data Fig. 5.

A limitation of the retrieval analysis is that the inferred $T$–$P$ profile is not derived self-consistently with the inferred chemical abundances. The latter determine the opacity of the atmosphere, which in turn controls the radiative transfer and thus the $T$–$P$ profile. As a check, we used the chemical abundances obtained from the retrieval analysis to generate self-consistent $T$–$P$ profiles in radiative-convective equilibrium. In one test, the VO abundance was set to the median value from the retrieval, and in a second test it was set to the lower value of the $1\sigma$ credible range. In both cases, the $H_2O$ abundance was set to its median value from the retrieval. The results are shown in Extended Data Fig. 4. Between pressures of about $10^{-3}$ bar to $10^{-5}$ bar, the temperatures of the radiative-convective equilibrium models range from approximately 2,400 K to 3,000 K, in agreement with the retrieved $T$–$P$ profile shown in Fig. 3 at the uncertainty levels. The thermal spectra show $H_2O$ and VO emission features, with overall flux levels broadly matching those measured in the WFC3 G141 bandpass but giving a poorer agreement for the 3.6 µm IRAC data point (Extended Data Fig. 4). The approximate nature of the parameterized $T$–$P$ profile is most prominently seen at the lowest pressures below ~$10^{-5}$ bar where it is forced to be isothermal at about 2,900 K (Fig. 3), whereas the radiative-convective models rise to temperatures exceeding 3,000 K. This results in isothermal blackbody emission at optical wavelengths (Fig. 2), in contrast to the self-consistent model which shows prominent optical VO emission features. Furthermore, the retrieved VO abundances are probably biased towards large values, as the isothermal approximation at low pressures limits the emission contributions from those pressure levels. However, the $H_2O$ retrieved abundance and emergent spectra are expected to be less affected, as the contribution functions at these wavelengths are deeper in the atmosphere, where the parameterized retrieval $T$–$P$ profile used for the retrieval is more accurate.

To help quantify the corresponding optical opacities needed to produce the observed stratosphere, we implemented the approach of refs 11 and 12, in which an arbitrary absorber was added throughout the atmosphere with grey opacity across the 0.43–1 µm optical wavelength range. We assumed 1× and 15× solar elemental abundances, with gaseous TiO and VO opacities removed. With this set-up, a grid of models in radiative-convective equilibrium was generated for a range of heat recirculation factors and assuming the grey absorber had opacity $\kappa$ of 0.002, 0.02, 0.04, 0.06, 0.08, 0.1, 0.2 and 2 $cm^2$ $g^{-1}$. All models exhibited stratospheres between pressures of $10^{-2}$ bar and $10^{-5}$ bar. Two examples with zero heat recirculation are plotted in Extended Data Fig. 4, one for $\kappa = 0.02$ $cm^2$ $g^{-1}$ and solar metallicity, and the other with $\kappa = 0.06$ $cm^2$ $g^{-1}$ and 15× solar metallicity. Both show $H_2O$ in emission and reproduce the overall level of thermal emission measured in the WFC3 G141 bandpass, while over-predicting the flux measured in the 3.6 µm IRAC bandpass.

**Validating the ATMO retrieval code**

We benchmarked our ATMO retrieval and $T$–$P$ profile parameterization code by applying it to the published thermal spectrum of WASP-43b[4,47,48]. Retrieval analyses performed on these data with the CHIMERA code[2,3] have previously indicated near-solar abundances of $H_2O$ along with a non-inverted $T$–$P$ profile for the planetary atmosphere. We performed a retrieval with the same data and fit parameters as the CHIMERA study published in ref. 48. The latter consisted of five parameters for the $T$–$P$ profile and abundances for $H_2O$, CO, $CO_2$, $CH_4$, $NH_3$ and HCN. For consistency, we turned off scattering in our retrieval analysis, as this was not included in the CHIMERA study. We also note that CHIMERA uses the HITRAN database[49] for many of the molecular opacities, including $H_2O$, whereas ATMO primarily uses the ExoMol database[50]. We find good agreement with the retrieved $T$–$P$ profiles and $H_2O$ abundance reported in ref. 48 (Extended Data Fig. 6).

**Code availability.** Custom code used to extract the HST spectra from the raw data frames and custom code used to extract and analyse the Spitzer data are available upon request. Publicly available custom codes were used for the Gaussian process modelling (http://github.com/tomevans/gps) and eclipse signal modelling (http://github.com/tomevans/planetc), both of which are publicly available but not maintained in a user-friendly state. The HST light curve MCMC fitting was performed using the open source emcee code (http://github.com/dfm/emcee). The MCMC retrieval analyses were performed using the publicly available package exofast (http://astroutils.astronomy.ohio-state.edu/exofast). The ATMO code used to compute the atmosphere models is currently proprietary.

**Data availability.** Raw HST data frames are publicly available online at the Mikulski Archive for Space Telescopes (MAST; https://archive.stsci.edu). Raw Spitzer data frames are publicly available online at the NASA/IPAC Infrared Science Archive (IRSA; http://irsa.ipac.caltech.edu/Missions/spitzer. html). Reduced data products and models used in this study are available in Supplementary Information.

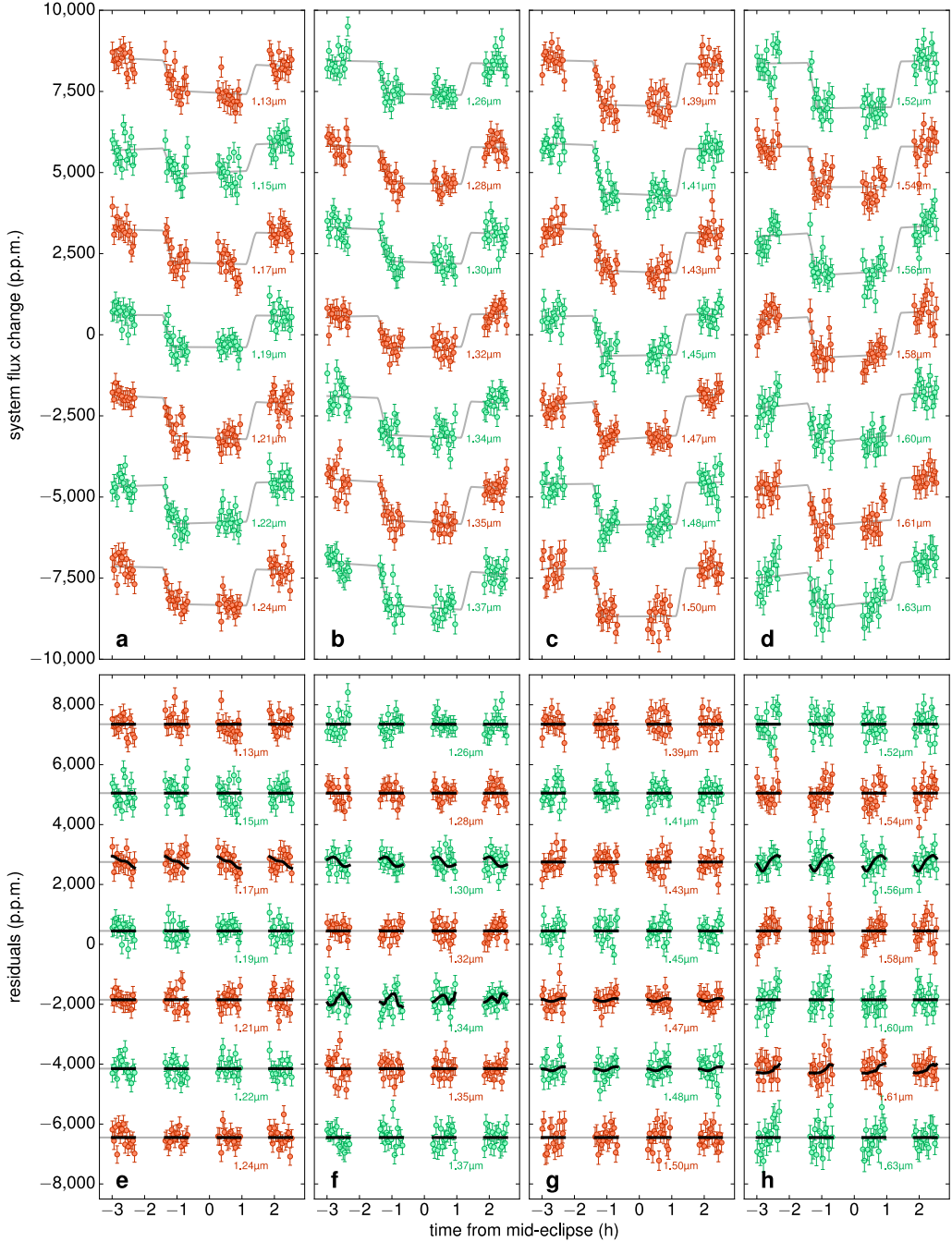

**Extended Data Figure 1 | Spectroscopic light curves for WASP-121b. a–d**, Raw normalized light curves for each of the spectroscopic channels with photon-noise $1\sigma$ error bars (orange and green circles), and best-fit eclipse signals multiplied by linear time trends (grey lines). Vertical offsets have been applied for visual clarity. Labels indicate central wavelengths for each channel. **e–h**, Model residuals after removing the best-fit eclipse signal and linear time trend for each of the spectroscopic light curves in **a–d**, respectively, with photon-noise $1\sigma$ error bars (orange and green circles) and stochastic components of the best-fit Gaussian process models (black lines). Note that for most channels, the correlations remaining in the residuals after accounting for a linear time trend are minimal.

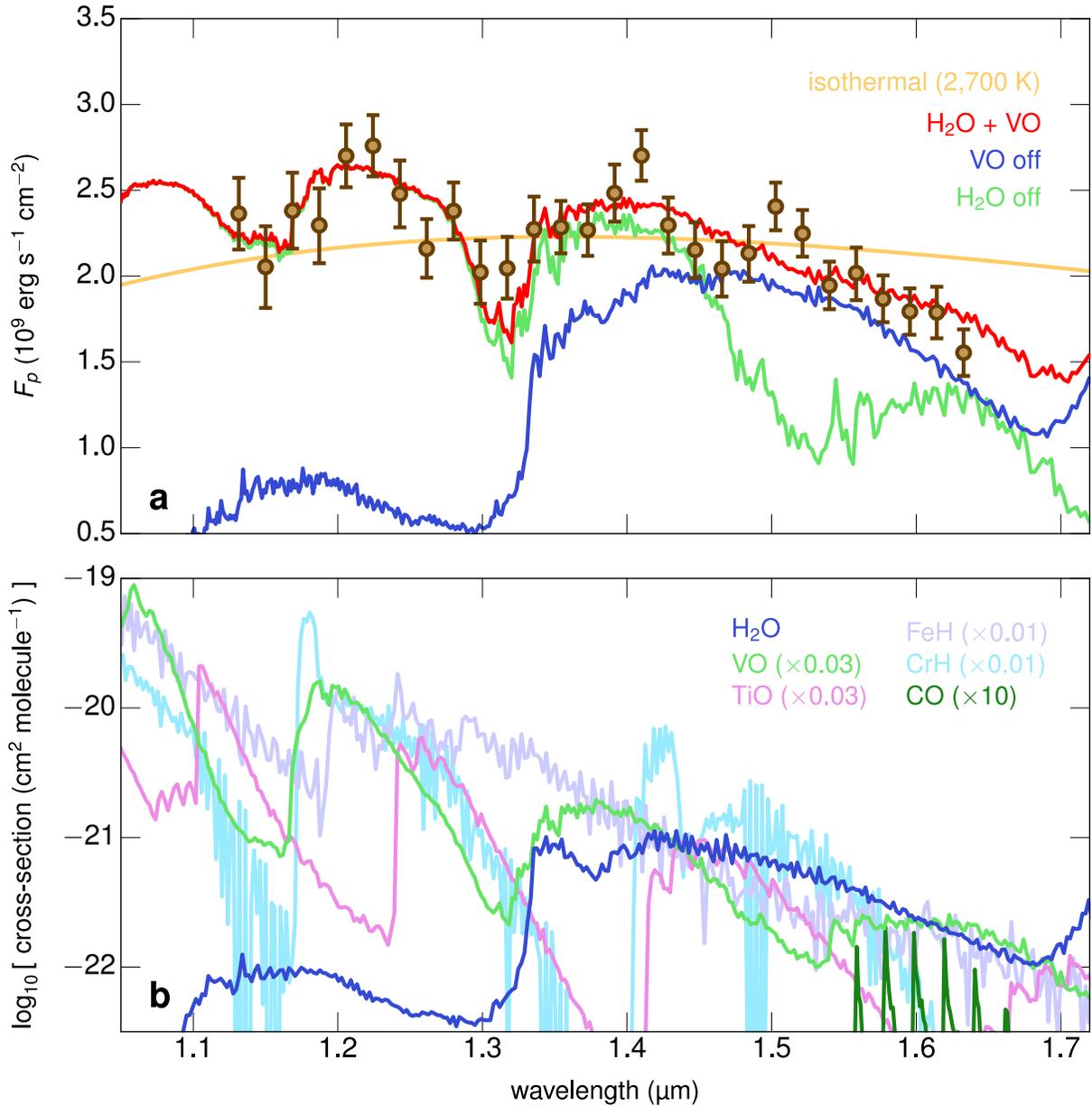

**Extended Data Figure 2 | Model thermal spectrum for WASP-121b broken down by emission source and absorption cross-sections of important molecules. a**, Similar to Fig. 2, showing the HST measurements of the WASP-121b thermal spectrum (brown circles; error bars, 1$\sigma$ uncertainties). Red line shows the best-fit model (H$_2$O and VO) obtained from the retrieval analysis. Other coloured lines show model thermal spectra generated using the same *T–P* profile as for the best-fit model but with the opacity due to each molecule turned off one-by-one: VO off (blue line) and H$_2$O off (green line). **b**, Absorption cross-sections for plausible gas-phase absorbers (colour coded, key at top right) across the WFC3 bandpass. Text labels in the key give the rescaling factors that have been applied to each cross-section to fit them on a single vertical axis, with the exception of the H$_2$O cross-section which has not been rescaled. Note that the VO cross-section has been rescaled by a factor of 0.03, which is consistent with the abundance of VO relative to H$_2$O inferred for the best-fit model (Extended Data Table 4).

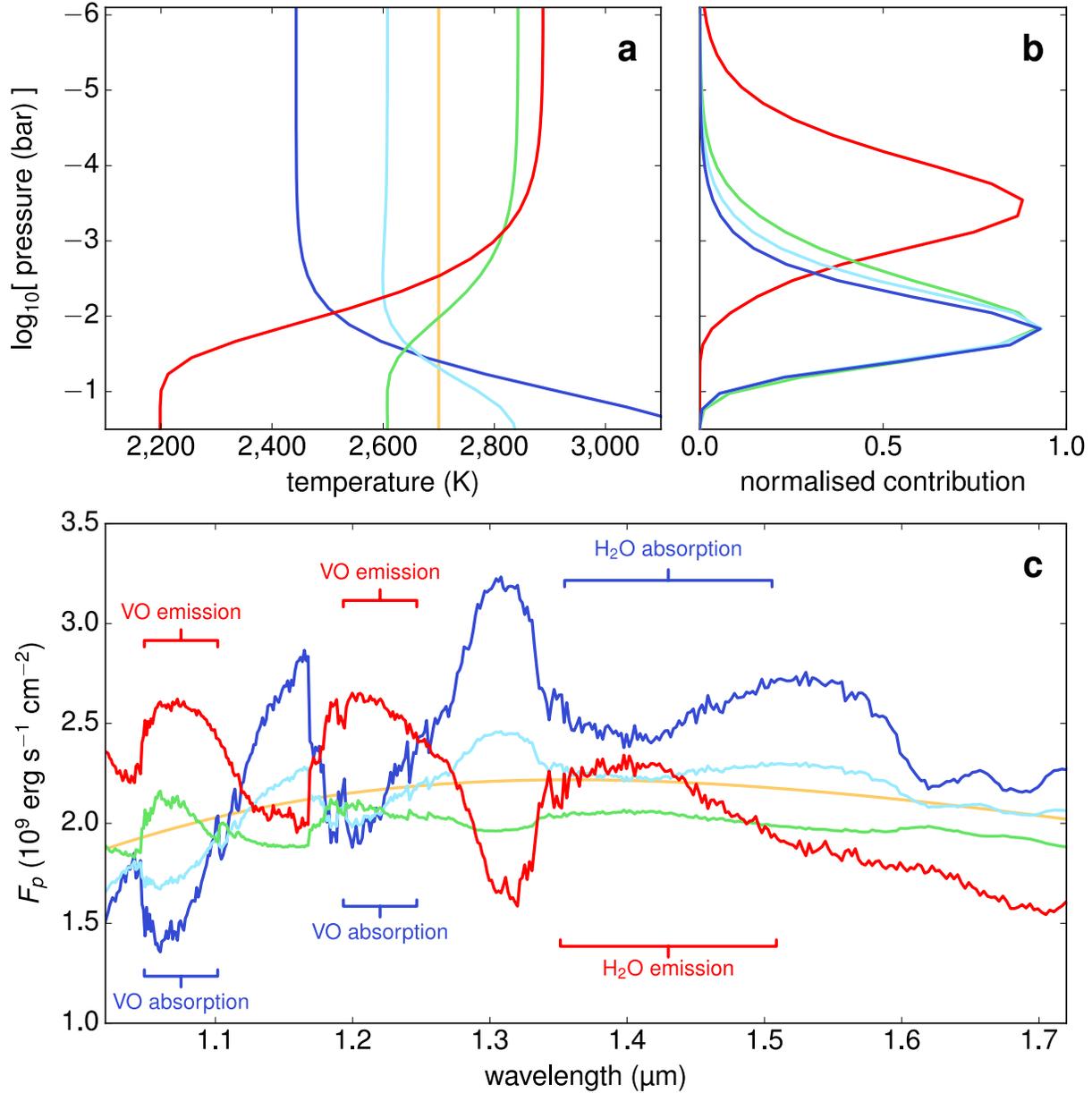

**Extended Data Figure 3 | A comparison of models with and without stratospheres. a**, A sequence of *T–P* profiles with and without stratospheres computed using the analytic parameterization of ref. 23. The red line corresponds to a model with a strong stratosphere obtained from the MCMC retrieval analysis and the green line shows a model with a weaker stratosphere. The light blue and dark blue lines show, respectively, models with a decreasing temperature profile and a strongly-decreasing temperature profile. The yellow line indicates the best-fit isothermal model with a temperature of 2,700 K. **b**, Corresponding contribution functions averaged over the WFC3 G141 bandpass, indicating the approximate pressures probed at these wavelengths. **c**, Corresponding thermal spectra, with $H_2O$ and VO bands seen in emission for the models with stratospheres (red and green lines) and in absorption for the models with decreasing temperature profiles (light blue and dark blue lines). The isothermal model has a featureless blackbody spectrum (yellow line).

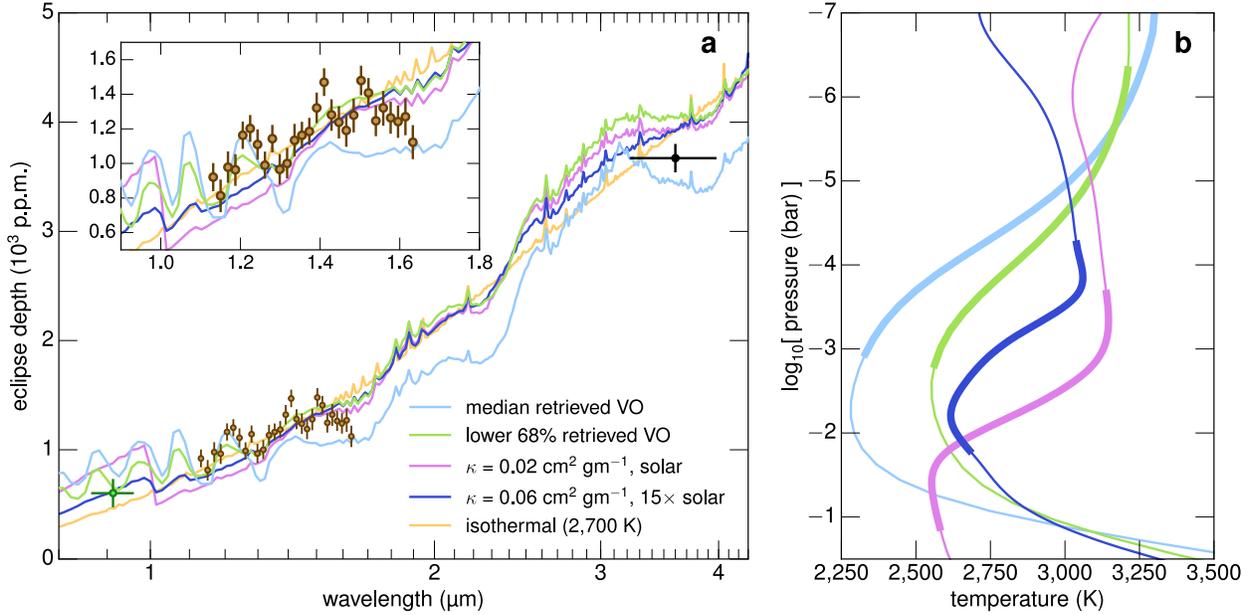

**Extended Data Figure 4 | Self-consistent models in radiative-convective equilibrium. a**, Similar to Fig. 2, but showing self-consistent models in radiative equilibrium: median $H_2O$ and VO abundances obtained from the retrieval analysis (light blue line); $H_2O$ abundance set to median value and VO abundance set to the lower value of the 68% credible range (green line); solar metallicity, but without VO and TiO, and with an arbitrary absorber with absorption cross-section $\kappa = 0.02$ cm$^2$ g$^{-1}$ (pink line); and 15× solar metallicity, but without VO and TiO, and with an arbitrary absorber with $\kappa = 0.06$ cm$^2$ g$^{-1}$ (dark blue line). For the last two models, the arbitrary absorber has grey opacity across the 0.43–1 μm wavelength range and is assumed to be distributed uniformly throughout the atmosphere. The best-fit isothermal model with temperature 2,700 K is also shown (yellow line). Brown circles with error bars, HST data with 1$\sigma$ uncertainties: inset, HST data on a magnified scale. **b**, Corresponding *T–P* profiles, all of which have a stratosphere across the $10^{-2}$ bar to $10^{-5}$ bar pressure range. Thick lines indicate the main pressure levels probed across the WFC3 G141 bandpass.

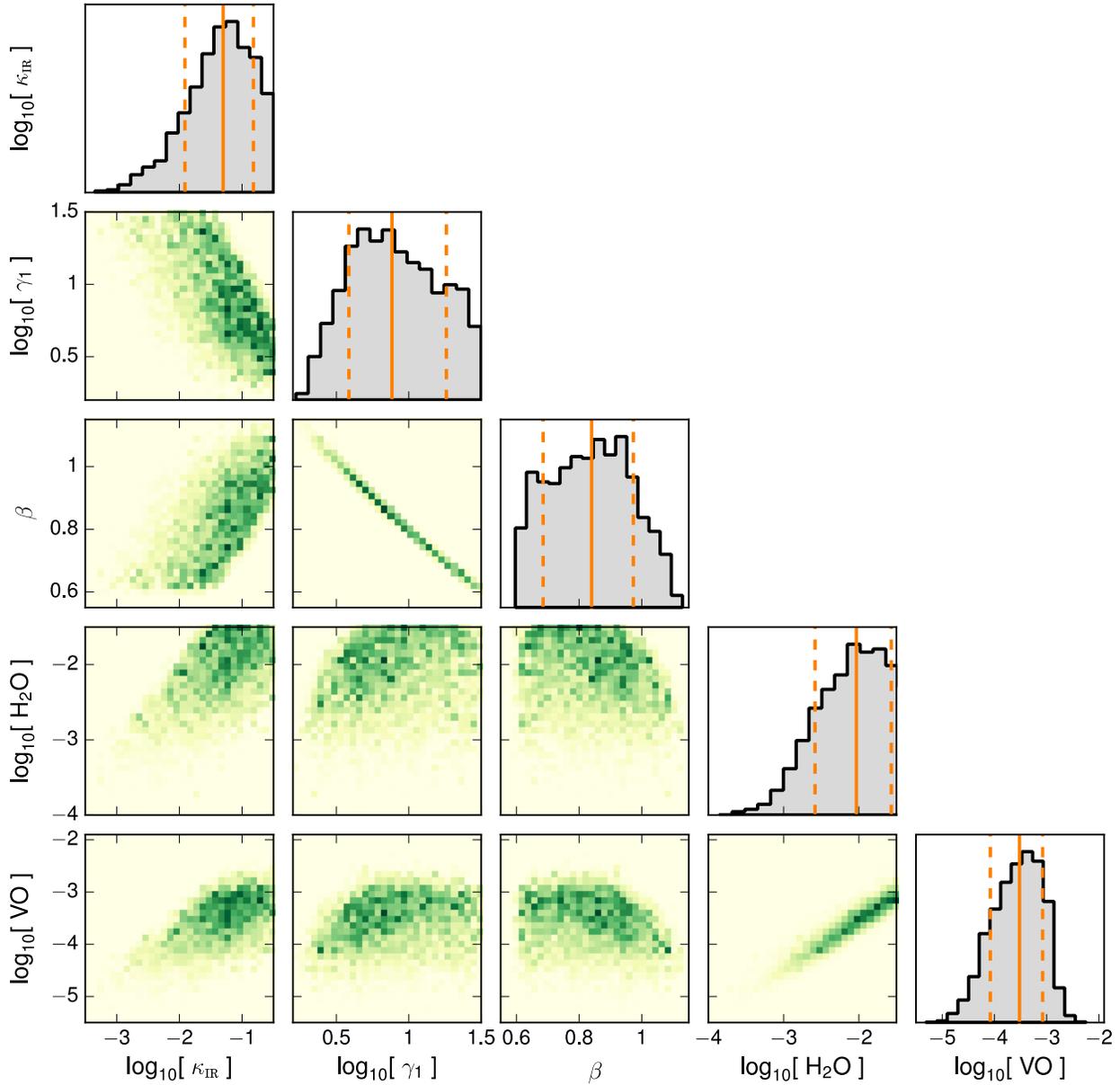

**Extended Data Figure 5 | Posterior distributions from MCMC retrieval analysis for WASP-121b.** Panels along the diagonal show the marginalized posterior distributions for each parameter of the thermal emission model ($\kappa_{IR}$, $\gamma_1$, $\beta$, $H_2O$ abundance, VO abundance), shown left to right in the columns, and top to bottom in the rows. Solid orange lines show the posterior medians and dashed orange lines show the ranges either side encompassing ±34% of the samples. Panels below the diagonal show the posterior distribution as a function of each parameter pair with increasing probability from light to dark shades. As $\gamma_1$ controls the optical opacity, there is a strong anti-correlation with the radiation efficiency factor $\beta$, where lower radiation efficiency values are compensated by higher optical opacities and vice versa. There is also a strong degeneracy between the abundances of each molecule ($H_2O$, VO) and the parameter $\kappa_{IR}$ which controls the overall atmospheric opacity. Higher abundances and higher values of $\kappa_{IR}$ result in stratospheres at lower pressures and vice versa.

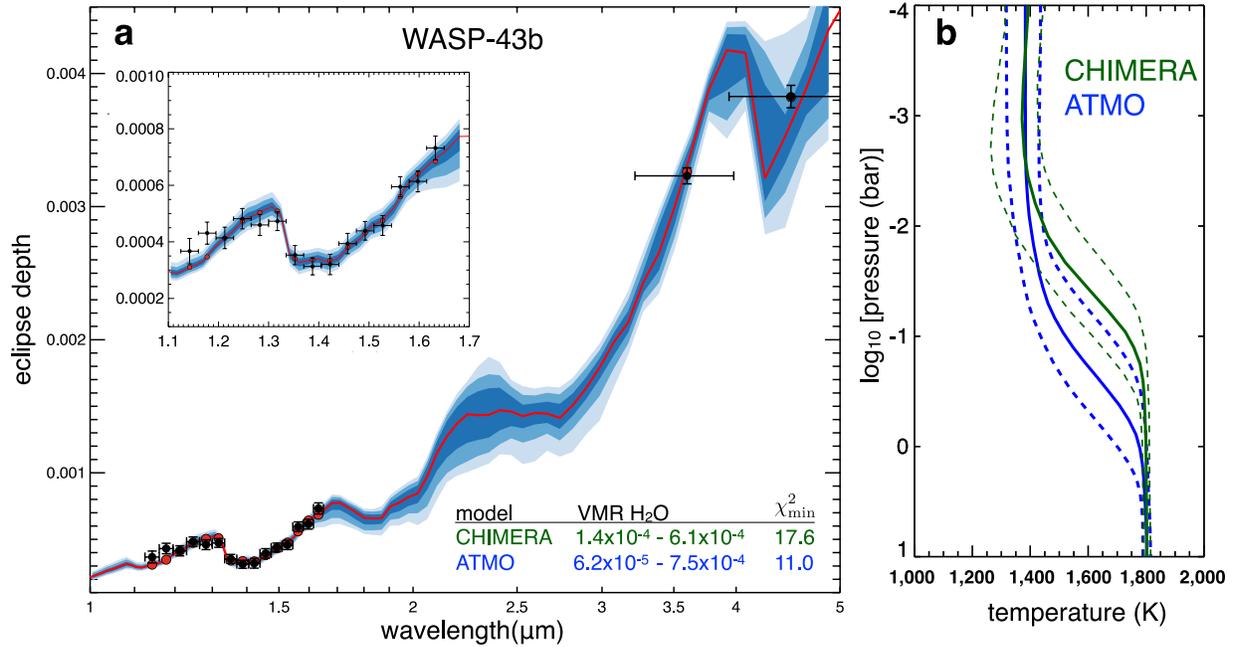

**Extended Data Figure 6 | ATMO retrieval code applied to WASP-43b thermal emission data. a,** Black data points show the measured thermal emission data for WASP-43b taken from ref. 48, composed of HST/WFC3 and Spitzer/IRAC observations. Vertical error bars give $1\sigma$ uncertainties and horizontal error bars give photometric bandpasses. The red line shows the best-fit spectrum obtained from an ATMO retrieval analysis, with shading indicating the regions encompassing 68%, 95% and 99.7% of the MCMC samples. The inset shows a magnified view of the HST data. The retrieved volume mixing ratio (VMR) for $H_2O$ is in good agreement with the published value obtained using the CHIMERA retrieval code[2,3] and the ATMO retrieval achieves a better fit to the data as measured by the minimum $\chi^2$. **b,** Best-fit $T$–$P$ profiles retrieved with ATMO (solid blue line) and CHIMERA (solid green line). The two analyses agree to within the 68% credible ranges (blue and green dashed lines).

**Extended Data Table 1 | Parameter values for the WASP-121b white light curve fit**

| Free parameters | Value | Units |
|---|---|---|
| eclipse depth | $1132^{+37}_{-35}$ | p.p.m. |
| $T_{mid}$ | $2457703.45879^{+0.00039}_{-0.00039}$ | $BJD_{TDB}$ |
| $c_0$ | $0.9996^{+0.0009}_{-0.0021}$ | — |
| $c_1$ | $-0.000217^{+0.000046}_{-0.000061}$ | arbitrary normalization |
| $A$ | $1431^{+3305}_{-748}$ | ppm |
| $\log[\eta_\phi]$ | $-0.35^{+1.14}_{-1.57}$ | arbitrary normalization |
| $\log[\eta_x]$ | $-2.87^{+1.81}_{-1.83}$ | arbitrary normalization |
| $\log[\eta_y]$ | $-2.82^{+2.78}_{-2.62}$ | arbitrary normalization |
| **Fixed parameters** | **Value** | **Units** |
| $P$ | 1.2749255 | day |
| $R_p$ | 1.694 | $R_J$ |
| $R_s$ | 1.458 | $R_\odot$ |
| $a/R_s$ | 3.825 | — |
| $i$ | 88.91 | degree |
| $b = a\cos(i)/R_s$ | 0.073 | — |
| $e$ | 0 | — |

Free parameters are those that were allowed to vary in the model fitting, and fixed values were obtained from previous analyses of primary transit light curves[19,20]. For the free parameters, quoted values give the median of the MCMC samples with ranges either side encompassing ±34% of the samples. Note that the eclipse depth is equal to $(R_p/R_s)^2(F_p/F_s)$. Values for $c_1$, $\log[\eta_\phi]$, $\log[\eta_x]$ and $\log[\eta_y]$ have arbitrary normalizations owing to standardization of their associated variables before fitting (see Methods). The eclipse mid-time $T_{mid}$ is quoted as a Barycentric Julian Date in the Barycentric Dynamical Time standard ($BJD_{TDB}$), which was determined by converting from $JD_{UTC}$ using the online tools of ref. 51. To do so, we assumed a geocentric reference frame for HST, which introduces an error of up to ~0.02 s. However, this is insignificant compared to the ~1 s uncertainty of the HST timestamps[51] and the ~30 s uncertainty in the measurement of $T_{mid}$ itself.

**Extended Data Table 2 | WFC3 spectroscopic eclipse depths for WASP-121b**

| Wavelength (μm) | Eclipse depth (p.p.m.) | | Wavelength (μm) | Eclipse depth (p.p.m.) |
|---|---|---|---|---|
| 1.122-1.141 | $921^{+81}_{-81}$ | | 1.382-1.401 | $1321^{+89}_{-88}$ |
| 1.141-1.159 | $813^{+94}_{-95}$ | | 1.401-1.419 | $1470^{+81}_{-80}$ |
| 1.159-1.178 | $978^{+91}_{-90}$ | | 1.419-1.438 | $1281^{+90}_{-92}$ |
| 1.178-1.196 | $963^{+90}_{-93}$ | | 1.438-1.456 | $1238^{+92}_{-94}$ |
| 1.196-1.215 | $1163^{+79}_{-79}$ | | 1.456-1.475 | $1191^{+94}_{-94}$ |
| 1.215-1.234 | $1202^{+78}_{-78}$ | | 1.475-1.494 | $1279^{+95}_{-100}$ |
| 1.234-1.252 | $1109^{+87}_{-87}$ | | 1.494-1.512 | $1479^{+86}_{-85}$ |
| 1.252-1.271 | $989^{+79}_{-78}$ | | 1.512-1.531 | $1408^{+86}_{-85}$ |
| 1.271-1.289 | $1143^{+80}_{-80}$ | | 1.531-1.549 | $1246^{+89}_{-89}$ |
| 1.289-1.308 | $966^{+88}_{-88}$ | | 1.549-1.568 | $1321^{+98}_{-103}$ |
| 1.308-1.326 | $1000^{+89}_{-86}$ | | 1.568-1.586 | $1264^{+93}_{-92}$ |
| 1.326-1.345 | $1133^{+95}_{-94}$ | | 1.586-1.605 | $1242^{+93}_{-94}$ |
| 1.345-1.364 | $1163^{+78}_{-78}$ | | 1.605-1.623 | $1270^{+106}_{-107}$ |
| 1.364-1.382 | $1185^{+78}_{-78}$ | | 1.623-1.642 | $1123^{+98}_{-98}$ |

Secondary eclipse depths for each of the WFC3 spectroscopic channels obtained from the MCMC light curve analyses. Quoted values give the median of the MCMC samples with ranges either side encompassing ±34% of the samples.

**Extended Data Table 3 | Comparison of different retrieval analyses for WASP-121b**

| | Molecules | $n$ | $\chi^2$ | B.I.C. | | Molecules | $n$ | $\chi^2$ | B.I.C. |
|---|---|---|---|---|---|---|---|---|---|
| 1 | $H_2O$, VO | 25 | 26.6 | 43.6 | 19 | VO, FeH, CrH | 24 | 30.3 | 50.7 |
| 2 | $H_2O$, VO, FeH | 24 | 23.7 | 44.1 | 20 | VO, TiO | 25 | 34.2 | 51.2 |
| 3 | $H_2O$, VO, CrH | 24 | 25.1 | 45.5 | 21 | VO, TiO, FeH | 24 | 31.9 | 52.3 |
| 4 | $H_2O$, VO, TiO | 24 | 25.5 | 45.9 | 22 | VO | 26 | 39.1 | 52.7 |
| 5 | $H_2O$, VO, TiO, FeH | 23 | 22.4 | 46.2 | 23 | VO, $CH_4$ | 25 | 38.7 | 55.7 |
| 6 | $H_2O$, VO, $CH_4$ | 24 | 26.3 | 46.7 | 24 | VO, FeH | 25 | 39.1 | 56.1 |
| 7 | $H_2O$, VO, CO | 24 | 26.5 | 46.9 | 25 | $H_2O$, CO | 25 | 44.6 | 61.6 |
| 8 | $H_2O$, VO, $NH_3$ | 24 | 26.5 | 46.9 | 26 | CrH | 26 | 49.0 | 62.6 |
| 9 | VO, CrH | 25 | 30.3 | 47.3 | 27 | $H_2O$ | 26 | 49.3 | 62.9 |
| 10 | $H_2O$, VO, FeH, $CH_4$ | 23 | 23.6 | 47.4 | 28 | $H_2O$, CrH | 25 | 48.0 | 65.0 |
| 11 | $H_2O$, VO, FeH, $NH_3$ | 23 | 23.7 | 47.5 | 29 | TiO, FeH | 25 | 48.6 | 65.6 |
| 12 | $H_2O$, VO, FeH, CO | 23 | 23.8 | 47.6 | 30 | $H_2O$, TiO | 25 | 50.0 | 67.0 |
| 13 | VO, TiO, CrH | 24 | 28.1 | 48.5 | 31 | TiO | 26 | 54.8 | 68.4 |
| 14 | $H_2O$, VO, TiO, CO | 23 | 25.3 | 49.2 | 32 | $H_2O$, TiO, FeH | 24 | 48.1 | 68.5 |
| 15 | $H_2O$, VO, TiO, FeH, CO | 22 | 22.5 | 49.7 | 33 | $H_2O$, TiO, CO | 24 | 50.0 | 70.4 |
| 16 | $H_2O$, VO, $CH_4$, $NH_3$ | 23 | 26.3 | 50.1 | 34 | $H_2O$, TiO, FeH, CO | 23 | 47.5 | 71.3 |
| 17 | VO, $NH_3$ | 25 | 33.3 | 50.3 | 35 | $H_2O$, FeH | 25 | 60.7 | 77.7 |
| 18 | $H_2O$, VO, FeH, $CH_4$, $NH_3$ | 22 | 23.5 | 50.7 | 36 | $H_2O$, FeH, CO | 24 | 63.9 | 84.3 |

Summary of the different models for which retrieval analyses were performed, where $n$ is the number of degrees of freedom. The quality of the obtained fits to the measured thermal spectrum are indicated by the $\chi^2$ and Bayesian information criterion (BIC) values. Models are arranged in order of increasing BIC value. Note that the best-fit isothermal model has $\chi^2 = 83.9$ for $n = 29$ (see main text), corresponding to a BIC value of 85.9.

**Extended Data Table 4 | MCMC retrieval analysis results for WASP-121b**

| Parameter | Value | Units |
|:---:|:---:|:---:|
| $\log_{10}[\ \gamma_1\ ]$ | $0.38^{+16}_{-13}$ | — |
| $\log_{10}[\ \kappa_{\mathrm{IR}}\ ]$ | $-0.57^{+0.21}_{-0.26}$ | $\log_{10}[\ \mathrm{cm}^2\,\mathrm{g}^{-1}\ ]$ |
| $\beta$ | $0.84^{+0.13}_{-0.15}$ | — |
| $\log_{10}[\ H_2O\ ]$ | $-2.0^{+0.5}_{-0.5}$ | — |
| $\log_{10}[\ VO\ ]$ | $-3.5^{+0.4}_{-0.6}$ | — |

Quoted values give the median of the MCMC samples with ranges either side encompassing ±34% of the samples.